\documentclass[twocolumn,
superscriptaddress,preprintnumbers,amsmath,amssymb,
]{revtex4}

\usepackage{kotex}

\usepackage{amssymb,amsmath,amsthm,amsfonts,amstext,amsbsy,amscd}
\usepackage{bbold}
\usepackage{graphicx,epsfig}
\usepackage{mathrsfs}
\usepackage{multirow}
\usepackage{xcolor}

\usepackage{datetime}

\newtheorem{Def}{Definition}
\newtheorem{Prop}{Proposition}

\newtheorem{Thm}{Theorem}
\newtheorem{Cor}{Corollary}
\newtheorem{Rem}{Remark}
\newtheorem{Example}{Example}

\newcommand{\ket}[1]{{| #1 \rangle}}
\newcommand{\bra}[1]{{\langle #1 |}}
\def\ketbra#1#2{{\vert#1\rangle\!\langle#2\vert}}

\def\Complex{\mathbb{C}}

\newcommand*{\cC}{\mathcal{C}}
\newcommand*{\cE}{\mathcal{E}}

\newcommand*{\cI}{\mathcal{I}}

\newcommand*{\cL}{\mathcal{L}}
\newcommand*{\cM}{\mathcal{M}}
\newcommand*{\cN}{\mathcal{N}}

\begin{document}

\title{
Sufficient conditions for additivity of the zero-error classical capacity of quantum channels
}
\author{Jeonghoon Park}
\affiliation{
Department of Applied Mathematics and Institute of Natural Sciences, 
Kyung Hee University, Yongin 17104, Korea
}
\author{Jeong San Kim}
\email{freddie1@khu.ac.kr}
\affiliation{
Department of Applied Mathematics and Institute of Natural Sciences, 
Kyung Hee University, Yongin 17104, Korea
}

\begin{abstract}
The one-shot zero-error classical capacity of a quantum channel is the maximum amount of classical information that can be transmitted with zero probability of error via a single channel use. This capacity is fundamentally characterized by the logarithm of the independence number of the noncommutative graph induced by the quantum channel. Consequently, the additivity of the one-shot zero-error classical capacity is equivalent to the multiplicativity of the independence number of the associated noncommutative graphs. As the independence number is not multiplicative in general, the specific conditions under which multiplicativity is preserved remain an open question in quantum information theory. Here, we establish some sufficient conditions for the multiplicativity of the independence number and provide explicit examples of quantum channels that satisfy these criteria. Furthermore, we investigate the block form of noncommutative graphs and derive the conditions under which the independence number remains multiplicative within this framework. These results offer new insights into the structural properties of noncommutative graphs and the fundamental limits of zero-error quantum communication.
\end{abstract}

\maketitle

\section{Introduction}
In information theory, the ordinary capacity of a communication channel is concerned with an error probability that tends to zero as the number of channel uses increases. In contrast, zero-error channel capacity is a more stringent measure, defined as the maximum asymptotic rate at which information can be transmitted with an absolute zero probability of error~\cite{Shan56}. This capacity is particularly vital in high-security or high-precision applications where no error tolerance is permitted, or in scenarios where only a limited number of channel uses are available~\cite{KO98}.
First introduced by Shannon~\cite{Shan56}, the study of zero-error capacity has been extensively explored in the classical realm~\cite{KO98}. Since its generalization to the quantum domain~\cite{MA05}, research has expanded significantly, spanning various types of information such as classical, private, and quantum information~\cite{Duan09,Shirokov15,LY16} as well as different shared resources~\cite{CLMW11,DSW13,DW16}.

A channel capacity is said to be \emph{additive} if the capacity of two combined channels equals the sum of their individual capacities. While computing zero-error capacities is generally an NP-hard problem~\cite{BS07}, additivity provides a critical simplification for evaluating joint channel performance. However, zero-error capacities are famously non-additive, in general, for both classical and quantum channels~\cite{Shan56, CCH11}. Although non-additivity can be viewed as an advantage implying that a joint channel can transmit more information than the sum of its parts, it also renders the capacity of combined channels mathematically intractable. Consequently, identifying the conditions under which additivity occurs is not only theoretically significant but also necessary for practical capacity estimation. Despite its importance, the specific circumstances that trigger additivity remain poorly understood.

One of the most effective frameworks for addressing this problem is the study of noncommutative graphs induced by quantum channels~\cite{DSW13}. Fundamental zero-error capacities can be reformulated using the structural properties of these graphs. Specifically, the one-shot zero-error classical capacity of a quantum channel is given by the logarithmic value of the independence number of its induced noncommutative graph~\cite{DSW13}. Under this framework, the additivity of the one-shot capacity is equivalent to the multiplicativity of the independence number of the associated graphs. 

Here we establish a rigorous theoretical framework that identifies when the independence number of noncommutative graphs exhibits multiplicativity, thereby resolving the additivity of zero-error capacities for specific classes of quantum channels. We first present sufficient conditions under which the independence number is multiplicative, ensuring the additivity of the one-shot zero-error classical capacity. These conditions are then shown to extend to the asymptotic zero-error classical capacity, providing a broader scope for our findings. Furthermore, we investigate the structural properties of noncommutative graphs in block form to derive specific criteria for multiplicativity within this architecture. 
This approach is based on the premise that 
when a noncommutative graph is decomposable, 
the constituent parts serve as essential building blocks
for understanding certain properties of the given noncommutative graph. 
Finally, we provide explicit examples of quantum channels to demonstrate the practical application of our additivity conditions and validate the theoretical results.

This paper is organized as follows. Section~\ref{sec:Pre} provides the necessary background on noncommutative graphs and zero-error classical capacity. In Section~\ref{sec:Add}, we present some sufficient conditions for the additivity of this capacity in one-shot and asymptotic settings. Section~\ref{sec:Block} investigates the block structure of noncommutative graphs and characterizes the additivity of the one-shot zero-error classical capacity therein. We further provide some explicit examples to illustrate our results. The paper concludes with a summary of results in Section~\ref{sec:Sum}.

\section{Preliminaries}\label{sec:Pre}

We denote by $\Complex^{n}$ the complex $n$-space. 
For two finite-dimensional Hilbert spaces $A$ and $B$, 
let $\cL(A,B)$ denote the space of linear operators from $A$ to $B$. 
When $A=B$, we denote $\cL(A,A)$ by $\cL(A)$. 
The identity operator on $A$ is denoted by $I_A$. 
When $A=\Complex^{n}$, we denote $I_A$ as $I_n$. 
Let $O_{AB}$ denote the zero operator in $\cL(A,B)$. 
The dimension of a vector space $V$ is denoted as $\dim(V)$. 
The logarithm is taken to base $2$. 

\begin{Def}[{\cite{DSW13}}]\label{def:ZECC}
For a quantum channel $\cN: \cL(A)\rightarrow\cL(B)$ 
with a set of Kraus operators $\{E_{i}: i=1,\dots,\mu\}$, 
the \emph{one-shot zero-error classical capacity} $\cC_0^{(1)}(\cN)$ is defined as 
\begin{equation}
\cC_0^{(1)}(\cN)=\log{\alpha(\cN)}, 
\end{equation}
where $\alpha(\cN)$ is the maximum number $\nu$ of states $\ket{\psi_{1}}, \dots, \ket{\psi_{\nu}}$ 
such that 
\begin{equation}\label{eq:orthog.condition}
\ketbra{\psi_s}{\psi_t}\perp\mathrm{span}\{E_{i}^{\dag}E_{j} : i,j=1,\dots,\mu\} 
\end{equation}
for all $s,t\in\{1,\dots,\nu\}$ with $s\ne{}t$. 
The (asymptotic) \emph{zero-error classical capacity} $\cC_0(\cN)$ is defined by 
\begin{equation}\label{eq:C_0}
\cC_0(\cN)=\lim_{k\rightarrow\infty}\frac{1}{k}\cC_0^{(1)}(\cN^{\otimes{}k}). 
\end{equation}
If there are no distinct states $\ket{\psi_s}$ and $\ket{\psi_t}$ 
satisfying Eq.~(\ref{eq:orthog.condition}), 
we define $\alpha(\cN)=1$ by convention. 
\end{Def}

\begin{Def}[{\cite{DSW13}}]
For a subspace $S\subseteq\cL(A)$, 
with $S^{\dag}:=\{M^{\dag}: M\in{}S\}$, 
we call $S$ a \emph{noncommutative graph} 
if $S^{\dag}=S$ and $I_{A}\in{S}$.
\label{def_non_grapk} 
\end{Def}

\begin{Rem}\label{Rem:ch-NG}
We observe that 
the induced subspace 
\begin{equation}\label{eq:NGofCh}
S(\cN)\equiv\mathrm{span}\{E_{i}^{\dag}E_{j} : i,j=1,\dots,\mu\}
\end{equation} 
in Eq.~(\ref{eq:orthog.condition}) 
uniquely determines $\alpha(\cN)$, 
and is clearly a noncommutative graph. 
Conversely, for any noncommutative graph $S\subseteq\cL(A)$, 
there exists a quantum channel (not necessarily unique) 
$\cN:\cL(A)\rightarrow\cL(B)$ such that $S=S(\cN)$~\emph{\cite{Duan09}}, 
that is, any noncommutative graph corresponds to some quantum channel. 
Furthermore, we note that 
\begin{equation}
S(\cN\otimes\cM)=S(\cN)\otimes{}S(\cM) 
\end{equation}
for any quantum channels $\cN$ and $\cM$. 
Therefore, 
the zero-error classical capacity is fundamentally characterized by the noncommutative graph. 
\end{Rem}

Due to Remark~\ref{Rem:ch-NG}, 
the capacities of the quantum channel $\cN$ in Definition~\ref{def:ZECC} can be defined in terms of the corresponding noncommutative graph 
as follows. 

\begin{Def}[{\cite{DSW13}}]\label{def:ZECC-NG}
For a noncommutative graph $S\subseteq\cL(A)$, 
$\alpha(S)$ is the maximum number $\nu$ of states $\ket{\psi_1}, \dots, \ket{\psi_\nu}$ 
such that 
\begin{equation}\label{eq:orthog.condition.NG}
\ketbra{\psi_{s}}{\psi_{t}}\perp{}S 
\end{equation}
for all $s,t\in\{1,\dots,\nu\}$ with $s\ne{}t$. 
We call $\alpha(S)$ the \emph{independence number} of $S$. 
Furthermore, 
\begin{eqnarray}
\cC_{0}^{(1)}(S)&=&\log\alpha(S), \\
\cC_0(S)&=&\lim_{k\rightarrow\infty}\frac{1}{k}\cC_0^{(1)}(S^{\otimes{}k}). 
\end{eqnarray}
\end{Def}

We note from Eq.~(\ref{eq:orthog.condition}) that 
for any noncommutative graph $S\subseteq\cL(A)$ and any unitary $U\in\cL(A)$, 
$\alpha(USU^{\dag})=\alpha(S)$. 
In other words, 
unitarily equivalent noncommutative graphs have the same independence number. 

\begin{Prop}[\cite{Duan09}]\label{prop:alpha=1}
Let $S$ be a noncommutative graph, 
then $\alpha(S)=1$ if and only if $S^{\perp}$ has no rank-one operator. 
\end{Prop}

\section{Additivity of zero-error classical capacity of a quantum channel}\label{sec:Add}
 
In this section, 
we present several sufficient conditions for the additivity of the zero-error classical capacity. 
We say that the one-shot zero-error classical capacity 
$\cC_{0}^{(1)}$ is \emph{additive} for quantum channels $\cN$ and $\cM$ if 
\begin{equation}\label{eq:1-shot_additive}
\cC_{0}^{(1)}(\cN\otimes\cM)=\cC_{0}^{(1)}(\cN)+\cC_{0}^{(1)}(\cM), 
\end{equation}
and the zero-error classical capacity $\cC_0$ is \emph{additive} for quantum channels $\cN$ and $\cM$ if 
\begin{equation}
\cC_0(\cN\otimes\cM)=\cC_0(\cN)+\cC_0(\cM). 
\end{equation}

It follows from Remark~\ref{Rem:ch-NG} and Definition~\ref{def:ZECC-NG} that 
the additivity of quantum channels 
can also be defined in terms of the corresponding noncommutative graphs: 

\begin{Def}
We say that $\cC_{0}^{(1)}$ is \emph{additive} for noncommutative graphs $S$ and $T$ if 
\begin{equation}
\cC_0^{(1)}(S\otimes{}T)=\cC_0^{(1)}(S)+\cC_0^{(1)}(T). 
\end{equation}
Equivalently, we say that the independence number $\alpha$ is 
\emph{multiplicative} for $S$ and $T$ if 
\begin{equation}
\alpha(S\otimes{}T)=\alpha(S)\alpha(T). 
\end{equation}
Similarly, we say that $\cC_0$ is \emph{additive} for $S$ and $T$ if 
\begin{equation}
\cC_0(S\otimes{}T)=\cC_0(S)+\cC_0(T). 
\end{equation}
\end{Def}

In the remainder of this paper, 
we use the term \emph{additivity} of $\cC_{0}^{(1)}$ and $\cC_{0}$ 
for quantum channels and noncommutative graphs interchangeably. 

For the case that 
the noncommutative graph of a quantum channel is the whole space $\cL(A)$, 
the channel maps all states to a constant state. 
The following proposition says that 
both $\cC_{0}^{(1)}$ and $\cC_{0}$ are additive for two quantum channels 
when one of the two channels has the noncommutative graph equal to the whole space. 

\begin{Prop}[\cite{PH18}]\label{Prop:fullyNoisy}
Let $S\subseteq\cL(A)$ and $T\subseteq\cL(B)$ be noncommutative graphs. 
If $S=\cL(A)$, 
then $\cC_{0}^{(1)}$ and $\cC_{0}$ are additive for $S$ and $T$. 
\end{Prop}

We now consider the situation when one of the two quantum channel is 
a noiseless classical channel $\cE:\cL(A)\rightarrow\cL(A)$ defined as 
\begin{equation}\label{eq:noiselss_c.ch}
\cE(\rho)=\sum_{t=0}^{\dim(A)-1}\ketbra{e_t}{e_t}\rho\ketbra{e_t}{e_t}, 
\end{equation}
where $\{\ket{e_t} : t=0,\dots,\dim(A)-1\}$ is an orthonormal basis for a Hilbert space $A$. 
The channel $\cE$ in Eq.~(\ref{eq:noiselss_c.ch}) preserves the diagonal elements of $\rho$ with respect to $\ket{e_t}$'s, 
while eliminating all off-diagonal elements. In other words, 
a noiseless classical channel can perfectly transmit classical information while inevitably destroying quantum coherence. 

For a noiseless classical channel $\cE$ in Eq.~(\ref{eq:noiselss_c.ch}), we note that the induced noncommutative graph $S(\cE)$ is
$\mathrm{span}\{\ketbra{e_t}{e_t} : t=0,\dots,\dim(A)-1\}$. 
The following theorem shows that both $\cC_{0}^{(1)}$ and $\cC_{0}$ are additive for a pair of quantum channels when one of the channels is a noiseless classical channel. 
\begin{Thm}\label{Thm:additive_noiseless}
Let $S\subseteq\cL(\Complex^m)$ and $T\subseteq\cL(\Complex^n)$ be noncommutative graphs. 
If $S=\mathrm{span}\{\ketbra{e_t}{e_t} : t=0,\dots,m-1\}$, 
where $\{\ket{e_t} : t=0,\dots,m-1\}$ is an orthonormal basis for $\Complex^m$, 
then $\cC_{0}^{(1)}$ and $\cC_{0}$ are additive for $S$ and $T$. 
\end{Thm}
\begin{proof}
We observe that $\ketbra{e_i}{e_j}\perp{}S$ for all $i,j\in\{0,\dots,m-1\}$ with $i\ne{}j$, 
and so $\alpha(S)={m}$. 

Let $\alpha(T)={l}$ for some $1\le{}l\le{}n$. 
Assume to the contrary that 
\begin{equation}
\alpha(S\otimes{}T)>\alpha(S)\alpha(T). 
\end{equation}
It follow from Eq.~(\ref{eq:orthog.condition.NG}) that 
there are $ml+1$ states 
\begin{equation}
\ket{\psi_1},\cdots,\ket{\psi_{ml+1}}\in\Complex^{m}\otimes\Complex^{n} 
\end{equation}
such that 
\begin{equation}\label{eq:OrthoCond-temp}
\ketbra{\psi_i}{\psi_j}\perp{}S\otimes{}T 
\end{equation}
for all $i,j\in\Lambda\equiv\{1,\dots,ml+1\}$ with $i\ne{}j$. 
Let $\ket{\psi_i}=\sum_{t=0}^{m-1}\ket{e_t}\ket{v_t^{(i)}}$, 
where $\ket{v_t^{(i)}}\in\Complex^{n}$, 
then Eq.~(\ref{eq:OrthoCond-temp}) is equivalent to 
\begin{equation}\label{eq:codewords_additive}
\ketbra{v_t^{(i)}}{v_t^{(j)}}\perp{T} 
\end{equation}
for any $t\in\{0,\dots,m-1\}$. 
For each $t\in\{0,\dots,m-1\}$, we define 
\begin{equation}
\mu_{t}\equiv\{i\in\Lambda : \ket{v_{t}^{(i)}}\ne0\}. 
\end{equation}
Since $\alpha(T)=l$, 
we have $|\mu_{t}|\le{}l$ for any $t\in\{0,\dots,m-1\}$. 
Thus, we can see that 
\begin{equation}
|\{i\in\Lambda : \ket{\psi_i}\ne0\}|=\left|\bigcup_{t=0}^{m-1}\mu_{t}\right|\le{}ml. 
\end{equation}
This implies that $\ket{\psi_i}=0$ for some $i\in\Lambda$, 
which is a contradiction. 
We conclude that $\alpha(S\otimes{}T)=\alpha(S)\alpha(T)$, 
that is, $\cC_{0}^{(1)}(S\otimes{}T)=\cC_{0}^{(1)}(S)+\cC_{0}^{(1)}(T)$. 

Finally, we obtain $\cC_{0}(S\otimes{T})=\cC_{0}(S)+\cC_{0}(T)$ by noting that 
\begin{equation}
\cC_{0}^{(1)}\left((S\otimes{T})^{\otimes{k}}\right)
=k\cdot\cC_{0}^{(1)}(S)+\cC_{0}^{(1)}(T^{\otimes{k}}). 
\end{equation}
\end{proof}

\begin{Example}
Let us consider a Weyl channel $\cN: \cL(\Complex^n)\rightarrow\cL(\Complex^n)$ defined as 
\begin{equation}
\cN(\rho)=\sum_{k=0}^{n-1}p_{k}Z_{k}\rho{}Z_{k}^{\dag}, 
\end{equation}
where $p_{k}>0$ with $\sum_{k=0}^{n-1}p_{k}=1$, 
$Z_{k}=\sum_{j=0}^{n-1}\omega^{jk}\ketbra{j}{j}$, 
and $\omega=e^{2\pi\mathrm{i}/n}$ is an $n$th root of unity. 
It is straightforward to show that 
\begin{eqnarray}
S(\cN)
&=&\mathrm{span}\{Z_{k} : k=0,\dots,n-1\} \nonumber \\
&=&\mathrm{span}\{\ketbra{i}{i}: i=0,\dots,n-1\}, 
\end{eqnarray}
which corresponds to $S$ in Theorem~\ref{Thm:additive_noiseless}. 
Thus, both $\cC_0^{(1)}$ and $\cC_0$ are additive 
for the channel $\cN$ and any quantum channel $\cM$. 
\end{Example}

The noiseless \emph{classical} channel in Eq.~(\ref{eq:noiselss_c.ch}) ensures the additivity of $\cC_{0}^{(1)}$ and $\cC_{0}$ as shown in Theorem~\ref{Thm:additive_noiseless}. This naturally leads to the question if an analogous additivity holds for
the noiseless \emph{quantum} channel, that is, the identity channel.
However, the use of the noiseless quantum channel does not guarantee 
the additivity of $\cC_{0}^{(1)}$ and $\cC_{0}$; 
there exists a quantum channel $\cN$ such that 
$\cC_{0}^{(1)}(\cI_{2}\otimes\cN)>\cC_{0}^{(1)}(\cI_{2})+\cC_{0}^{(1)}(\cN)$, 
where $\cI_{2}$ is the noiseless qubit channel, 
that is, the identity channel on a two-dimensional quantum system~\cite{Duan09,PH18}. 

Now, let us consider a qubit channel $\cN:\cL(\Complex^{2})\rightarrow\cL(\Complex^{2})$. If $\cN$ is the noiseless qubit channel, 
then the corresponding noncommutative graph $S(\cN)$ equals 
$\Complex{}I_2:=\mathrm{span}\{cI_2: c\in\Complex\}$. 
In the following theorem, we show that 
for a qubit channel $\cN$ and any quantum channel $\cM$, 
if $\cN$ is noisy, 
that is, $S(\cN)\ne\Complex{}I_2$, 
then both $\cC_{0}^{(1)}$ and $\cC_{0}$ are additive 
for $\cN$ and $\cM$. 
\begin{Thm}\label{Thm:additivity_qubit}
Let $S\subseteq\cL(\Complex^{2})$ and $T\subseteq\cL(\Complex^{n})$ 
be noncommutative graphs. 
If $S\ne\Complex{}I_2$, 
then $\cC_{0}^{(1)}$ and $\cC_{0}$ are additive for $S$ and $T$. 
\end{Thm}
\begin{proof}
We note that any noncommutative graph $S\subseteq\cL(\Complex^2)$ is 
one of the following forms~\cite{DSW13}: 
\begin{equation}
\Complex{}I_2,~\mathrm{span}\{I_2, Z\},~\mathrm{span}\{I_2, Z, X\},~\cL(\Complex^2) 
\end{equation}
up to unitary equivalence, 
where $Z=\ketbra{0}{0}-\ketbra{1}{1}$ and $X=\ketbra{0}{1}+\ketbra{1}{0}$. 

We first suppose $S=\mathrm{span}\{I_2, Z\}$. 
Since 
\begin{equation}
\mathrm{span}\{I_2, Z\}=\mathrm{span}\{\ketbra{0}{0}, \ketbra{1}{1}\}, 
\end{equation}
by Theorem~\ref{Thm:additive_noiseless}, 
$\cC_0^{(1)}$ and $\cC_0$ are additive for $S$ and $T$. 

Suppose $S=\mathrm{span}\{I_2, Z, X\}$. 
If $\alpha(S\otimes{}T)=1$, then clearly $\alpha(S\otimes{}T)=\alpha(S)\alpha(T)$. 
We now assume $\alpha(S\otimes{}T)>1$. 
Thus, there exist $\alpha(S\otimes{}T)$ states 
\begin{equation}
\ket{\psi_1},\dots,\ket{\psi_{\alpha(S\otimes{}T)}}\in\Complex^2\otimes\Complex^{n} 
\end{equation}
satisfying 
\begin{equation}\label{eq:perp_temp1}
\bra{\psi_i}(S\otimes{}T)\ket{\psi_j}=0 
\end{equation}
for any $i,j\in\{1,\dots,\alpha(S\otimes{}T)\}$ with $i\ne{j}$. 
Let $\ket{\psi_i}=\ket{0}\ket{u_i}+\ket{1}\ket{v_i}$, 
where $\ket{u_i},\ket{v_i}\in\Complex^{n}$. 
By choosing $I_2,Z,X\in{}S$, from Eq.~(\ref{eq:perp_temp1}), 
we obtain 
\begin{equation}\label{eq:perp_temp2}
\ketbra{u_i}{u_j},~\ketbra{v_i}{v_j},~\ketbra{u_i}{v_j}+\ketbra{v_i}{u_j}\in{}T^{\perp}. 
\end{equation}
For each $i\in\{1,\dots,\alpha(S\otimes{}T)\}$, 
we let 
\begin{equation}
\ket{w_i}\equiv\left\{
\begin{array}{ll}
\ket{u_i} & \text{if}~\ket{u_i}\ne0, \\
\ket{v_i} & \text{if}~\ket{u_i}=0.
\end{array} \right.
\end{equation}
It follows from Eq.~(\ref{eq:perp_temp2}) that 
\begin{equation}
\ketbra{w_i}{w_j}\perp{}T 
\end{equation}
for any $i,j\in\{1,\dots,\alpha(S\otimes{}T)\}$ with $i\ne{j}$. 
Thus, $\alpha(S\otimes{}T)\le\alpha(T)$, 
and hence $\cC_0^{(1)}(S\otimes{}T)=\cC_0^{(1)}(S)+\cC_0^{(1)}(T)$. 
We then obtain from the observation 
\begin{equation}
\cC_0^{(1)}\left((S\otimes{}T)^{\otimes{}k}\right)
=k\cdot\cC_0^{(1)}(S)+\cC_0^{(1)}\left(T^{\otimes{}k}\right) 
\end{equation}
that $\cC_0(S\otimes{}T)=\cC_0(S)+\cC_0(T)$. 

Finally, the case of $S=\cL(\Complex^2)$ was proved in Proposition~\ref{Prop:fullyNoisy}. 
\end{proof}

We note that Theorem~\ref{Thm:additivity_qubit} generalizes the result of Theorem~\ref{Thm:additive_noiseless} for the case when $m=2$. Moreover, Theorem~\ref{Thm:additivity_qubit} also leads us to the following additivity property for the capacities of multi-qubit channels.

\begin{Cor}
For noncommutative graphs $S_{1},\dots,S_{n}\subseteq\cL(\Complex^{2})$, 
$\cC_0^{(1)}$ and $\cC_0$ are additive, 
that is, 
\begin{eqnarray}
\cC_0^{(1)}\left(\bigotimes_{i=1}^{n}S_{i}\right)&=&\sum_{i=1}^{n}\cC_0^{(1)}(S_{i}), \\
\cC_0\left(\bigotimes_{i=1}^{n}S_{i}\right)&=&\sum_{i=1}^{n}\cC_0(S_{i}). 
\end{eqnarray}
\end{Cor}
\begin{proof}
Without loss of generality, we assume that there exists $k$ such that 
$S_{i}=\Complex{}I_2$ for $i=1,\dots,k$, and $S_{i}\ne\Complex{}I_2$ for $i=k+1,\dots,n$ 
(up to unitary equivalence). Thus, we have 
\begin{eqnarray}
\cC_0^{(1)}\left(\bigotimes_{i=1}^{n}S_{i}\right)
&=&\cC_0^{(1)}\left({}(\Complex{}I_2)^{\otimes{}k}{}\right)+\sum_{i=k+1}^{n}\cC_0^{(1)}(S_{i}) \qquad \\
&=&\sum_{i=1}^{k}\cC_0^{(1)}(\Complex{}I_2)+\sum_{i=k+1}^{n}\cC_0^{(1)}(S_{i}), 
\end{eqnarray}
where the first equality is from Theorem~\ref{Thm:additivity_qubit}, 
and the second equality trivially holds. 
The additivity of $\cC_0$ is proved by the same argument. 
\end{proof}

\begin{Example}
Let us consider a quantum channel 
$\cN: \cL(\Complex^2)\rightarrow\cL(\Complex^2\otimes\Complex^2)$ defined as 
\begin{eqnarray}\label{eq:ch_Eg2}
\cN(\rho)
&=&\frac{1}{2}\left(\frac{1}{2}\rho+\frac{1}{2}Z\rho{}Z\right)\otimes\ketbra{0}{0} \nonumber \\
&{}&+\frac{1}{2}\left(\frac{1}{2}\rho+\frac{1}{2}X\rho{}X\right)\otimes\ketbra{1}{1}, 
\end{eqnarray}
where $Z=\ketbra{0}{0}-\ketbra{1}{1}$ and $X=\ketbra{0}{1}+\ketbra{1}{0}$. 
We note that the channel $\cN$ can be viewed as 
a mixture of a dephasing and a bit flip channels, 
which correspond to the first and second terms of the right-hand side of Eq.~(\ref{eq:ch_Eg2}), 
respectively. 
We can choose a set $\{E_{i}: i=0,1,2,3\}$ of Kraus operators for $\cN$, where 
\begin{eqnarray}
E_{0}=\frac{1}{2}(I_{2}\otimes\ket{0}),~E_{1}=\frac{1}{2}(Z\otimes\ket{0}), \nonumber \\
E_{2}=\frac{1}{2}(I_{2}\otimes\ket{1}),~E_{3}=\frac{1}{2}(X\otimes\ket{1}), 
\end{eqnarray}
and we can see that 
\begin{equation}
S(\cN)=\mathrm{span}\{I_{2}, Z, X\}. 
\end{equation}
It follows from Theorem~\ref{Thm:additivity_qubit} that 
$\cC_0^{(1)}$ and $\cC_0$ are additive for the channel $\cN$ and any quantum channel $\cM$. 
\end{Example}

\section{Block noncommutative graphs and additivity of zero-error classical capacity}\label{sec:Block}

Computing zero-error classical capacity of a noncommutative graph is generally a challenging task. 
Nevertheless, when a noncommutative graph can be decomposed into smaller components, the capacity could be evaluated from these constituent parts. 
On the other hand, by utilizing well-known noncommutative graphs as building blocks, we may construct a larger noncommutative graph that exhibits certain desired properties. 

In this section, we explore a block-structured form of noncommutative graphs and investigate its structural properties. We then provide specific conditions within this framework under which the one-shot zero-error classical capacity is additive.
\subsection{Block noncommutative graphs}

Let us consider the (external) direct sum $A\oplus{}B$ of Hilbert spaces $A$ and $B$, 
and a noncommutative graph $\Sigma\subseteq\cL(A\oplus{}B)$ of the form 
\begin{equation}\label{eq:BNG}
\Sigma=S\oplus{}T\oplus{}U\oplus{}V, 
\end{equation}
where 
\begin{equation}
S\subseteq\cL(A),~T\subseteq\cL(B),~U\subseteq\cL(B, A),~V\subseteq\cL(A, B) 
\end{equation}
are subspaces. 
We denote an element $M\in\cL(A\oplus{}B)$ by 
$M=(M_1, M_2, M_3, M_4)$, 
where $M_1\in\cL(A)$, $M_2\in\cL(B)$, $M_3\in\cL(B, A)$, and $M_4\in\cL(A, B)$. Because $\Sigma$ is a noncommutative graph, Definition~\ref{def_non_grapk} leads us to
$(I_{A}, I_{B}, O_{BA}, O_{AB})=I_{A\oplus{}B}\in\Sigma$ and 
$\Sigma^{\dag}=\Sigma$. In other words, $S$ and $T$ are also noncommutative graphs, and $V=U^{\dag}$, 
where $U^{\dag}=\{M^{\dag}: M\in{}U\}$. 

Conversely, let us consider any subspace  $\Sigma\subseteq\cL(A\oplus{}B)$ with the decomposition in Eq.~(\ref{eq:BNG}).
If $S$ and $T$ are noncommutative graphs and $V=U^{\dag}$, it naturally implies that $\Sigma^{\dag}=\Sigma$ and $I_{A\oplus{}B}=(I_{A}, I_{B}, O_{BA}, O_{AB})\in\Sigma$, therefore $\Sigma$ is a noncommutative graph. 
By denoting a subset $S$ of $\cL(X,Y)$ as $S_{XY}$, we have the  following definition of block noncommutative graph.
\begin{Def}
$\Sigma\subseteq\cL(A\oplus{}B)$ is called a \emph{block noncommutative graph} 
if there exist noncommutative graphs $S_{AA}$ and $T_{BB}$, and a subspace $U_{BA}$ 
such that 
\begin{equation}
\Sigma=S_{AA}\oplus{}T_{BB}\oplus{}U_{BA}\oplus{}U^{\dag}_{AB}. 
\end{equation}
\end{Def}

Now we are ready to have the following theorem stating that 
the independence number of a block noncommutative graph $\Sigma$ depends on all the direct summands of $\Sigma$. 
\begin{Thm}\label{Thm:blockNGcap}
For a block noncommutative graph 
\begin{equation}\label{eq:BNGform}
\Sigma=S_{AA}\oplus{}T_{BB}\oplus{}U_{BA}\oplus{}U^{\dag}_{AB}\subseteq\cL(A\oplus{}B), 
\end{equation}
the independence number $\alpha(\Sigma)$ is the maximum number of 
\begin{equation}
|\{\ket{v_i}\in{}A : i=1,\dots,m\}|+|\{\ket{w_s}\in{}B: s=1,\dots,n\}|, \quad
\end{equation}
where $\ket{v_i}$ and $\ket{w_s}$ are nonzero vectors 
such that 
\begin{eqnarray}
\ketbra{v_i}{v_j}&\perp&{}S_{AA},~\forall{}i,j\in\{1,\dots,m\}~\text{with}~i\ne{}j, \label{eq:perp_cond_1} \\
\ketbra{w_s}{w_t}&\perp&{}T_{BB},~\forall{}s,t\in\{1,\dots,n\}~\text{with}~s\ne{}t, \label{eq:perp_cond_2} \\
\ketbra{v_p}{w_q}&\perp&{}U_{BA},~\forall{}p\in\{1,\dots,m\},q\in\{1,\dots,n\}. \qquad \label{eq:perp_cond_3}
\end{eqnarray}
\end{Thm}
\begin{proof}
Assume that 
there exist $k$ states $\ket{\psi_1},\dots,\ket{\psi_k}\in{}A\oplus{}B$ 
satisfying Eq.~(\ref{eq:orthog.condition.NG}) with respect to $\Sigma$, 
that is, 
\begin{equation}\label{eq:orthog.Sigma}
\bra{\psi_i}\Sigma\ket{\psi_j}=0, 
\end{equation}
for all $i,j\in\{1,\dots,k\}$ with $i\ne{}j$. 
Let us write $\ket{\psi_i}\in{}A\oplus{}B$ as 
\begin{equation}
\ket{\psi_i}=(\ket{v_i}, \ket{w_i}), 
\end{equation}
where $\ket{v_i}\in{}A$ and $\ket{w_i}\in{}B$, 
then Eq.~(\ref{eq:orthog.Sigma}) is equivalent to
\begin{equation}\label{eq:orthog.Sigma-parts}
\bra{v_i}S_{AA}\ket{v_j}=\bra{w_i}T_{BB}\ket{w_j}=\bra{v_i}U_{BA}\ket{w_j}=0. 
\end{equation}
Suppose $\ket{v_t}$ and $\ket{w_t}$ are nonzero for some $t\in\{1,\dots,k\}$. 
We can see that Eq.~(\ref{eq:orthog.Sigma-parts}) still holds 
even if $\ket{\psi_t}$ is replaced by $(\ket{v_t}, 0)$ or $(0, \ket{w_t})$, 
where $0$ is the zero vector. 
Thus, it suffices to consider the cases when $\ket{\psi_i}=(\ket{v_i}, 0)$ 
and $\ket{\psi_i}=(0, \ket{w_i})$. 
We conclude from Eq.~(\ref{eq:orthog.Sigma-parts}) that $\alpha(\Sigma)$ is 
the maximum number of nonzero vectors $\ket{v_i}\in{}A$ and $\ket{w_s}\in{}B$ 
satisfying Eqs.~(\ref{eq:perp_cond_1})--(\ref{eq:perp_cond_3}). 
\end{proof}

The condition in Eq.~(\ref{eq:perp_cond_3}) implies that 
the subspace $U_{BA}^{\perp}$ has a rank-one operator $\ketbra{v_p}{w_q}$. 
Thus, if $U_{BA}^{\perp}$ has no rank-one operator, 
then exactly one of Eqs.~(\ref{eq:perp_cond_1}) and~(\ref{eq:perp_cond_2}) must hold. 
Furthermore, by applying the definition of the independence number $\alpha$ 
to Eqs.~(\ref{eq:perp_cond_1}) and~(\ref{eq:perp_cond_2}), 
we obtain the following corollary. 

\begin{Cor}\label{cor:offnorank1}
For a block noncommutative graph 
$\Sigma=S_{AA}\oplus{}T_{BB}\oplus{}U_{BA}\oplus{}U^{\dag}_{AB}\subseteq\cL(A\oplus{}B)$, 
\begin{equation}\label{eq:BNG-bounds}
\max\{\alpha(S_{AA}), \alpha(T_{BB})\}\le\alpha(\Sigma)\le\alpha(S_{AA})+\alpha(T_{BB}). \qquad
\end{equation}
In particular, if $U_{BA}^{\perp}$ has no rank-one operator, 
then $\alpha(\Sigma)=\max\{\alpha(S_{AA}), \alpha(T_{BB})\}$. 
\end{Cor}

The lower and upper bounds of $\alpha(\Sigma)$ in Eq.~(\ref{eq:BNG-bounds}) can be attained 
by choosing $U_{BA}=\cL(B,A)$ and $U_{BA}=\{O_{BA}\}$, respectively~\cite{DSW13}.

\subsection{Conditions on additivity of zero-error classical capacities 
for block noncommutative graphs}

We now present sufficient conditions for the additivity of one-shot zero-error classical capacity 
of block noncommutative graphs. 

\begin{Thm}~\label{Thm:add_BNG}
Let 
\begin{equation}
\Sigma=S_{AA}\oplus{}T_{BB}\oplus{}U_{BA}\oplus{}U^{\dag}_{AB}\subseteq\cL(A\oplus{}B) 
\end{equation}
be a block noncommutative graph and $\Omega\subseteq\cL(C)$ be a noncommutative graph. 
Suppose $\alpha$ is multiplicative for $S_{AA}$ and $\Omega$, 
and also for $T_{BB}$ and $\Omega$. 
It follows that $\alpha$ is multiplicative for $\Sigma$ and $\Omega$ 
if one of the following two conditions holds: 
\begin{eqnarray}
&(i)&\alpha(\Sigma)=\alpha(S_{AA})+\alpha(T_{BB}), \label{eq:sumcap} \\
&(ii)&U_{BA}=\cL(B,A)~\text{and}~\alpha(\Omega)=1. \label{eq:fulloff-0Omega}
\end{eqnarray}
\end{Thm}
\begin{proof}
First, we observe that 
\begin{equation}
\Sigma\otimes\Omega=(S\otimes\Omega)\oplus{}(T\otimes\Omega)\oplus{}(U\otimes\Omega)\oplus{}(U^{\dag}\otimes\Omega), 
\end{equation}
where $S\otimes\Omega\subseteq\cL(A\otimes{}C)$, 
$T\otimes\Omega\subseteq\cL(B\otimes{}C)$, 
$U\otimes\Omega\subseteq\cL(B\otimes{}C, A\otimes{}C)$, 
and $U^{\dag}\otimes\Omega\subseteq\cL(A\otimes{}C, B\otimes{}C)$. 
It is clear that $\Sigma\otimes\Omega\subseteq\cL((A\otimes{}C)\oplus(B\otimes{}C))$ 
is a block noncommutative graph. 

(i) Let us suppose $\alpha(\Sigma)=\alpha(S_{AA})+\alpha(T_{BB})$. 
It follows that 
\begin{eqnarray}
\alpha(\Sigma\otimes\Omega)
&\le&\alpha(S_{AA}\otimes\Omega)+\alpha(T_{BB}\otimes\Omega) \\
&=&\left(\alpha(S_{AA})+\alpha(T_{BB})\right)\alpha(\Omega) \\
&=&\alpha(\Sigma)\alpha(\Omega), 
\end{eqnarray}
where the inequality follows from Corollary~\ref{cor:offnorank1}, 
the first equality follows from the multiplicativity assumption for $\alpha$, 
and the second equality is from the assumption that 
$\alpha(\Sigma)=\alpha(S_{AA})+\alpha(T_{BB})$. 
On the other hand, 
$\alpha(\Sigma\otimes\Omega)\ge\alpha(\Sigma)\alpha(\Omega)$ holds trivially, 
and we conclude that $\alpha(\Sigma\otimes\Omega)=\alpha(\Sigma)\alpha(\Omega)$. 

(ii) We now suppose $U_{BA}=\cL(B,A)$ and $\alpha(\Omega)=1$. 
We first show by contradiction that $(\cL(B,A)\otimes\Omega)^{\perp}$ has no rank-one operator. 
Let us consider a block noncommutative graph 
\begin{eqnarray}
\cL(A\oplus{}B)\otimes\Omega
&=&(\cL(A)\otimes\Omega)\oplus{}(\cL(B)\otimes\Omega) \nonumber \\
&{}&\oplus{}(\cL(B,A)\otimes\Omega)\oplus{}(\cL(A,B)\otimes\Omega), \qquad
\end{eqnarray}
where $\cL(A)\otimes\Omega\subseteq\cL(A\otimes{}C)$, 
$\cL(B)\otimes\Omega\subseteq\cL(B\otimes{}C)$, 
$\cL(B,A)\otimes\Omega\subseteq\cL(B\otimes{}C, A\otimes{}C)$, 
and $\cL(A,B)\otimes\Omega\subseteq\cL(A\otimes{}C, B\otimes{}C)$. 
Assume that $(\cL(B,A)\otimes\Omega)^{\perp}$ has a rank-one operator. 
There exist two nonzero vectors $\ket{v}\in{}A\otimes{}C$ and $\ket{w}\in{}B\otimes{}C$ 
such that 
\begin{equation}
\ketbra{v}{w}\in(\cL(B,A)\otimes\Omega)^{\perp}. 
\end{equation} 
We can see that 
$\ket{v}$ and $\ket{w}$ satisfy Eqs.~(\ref{eq:perp_cond_1})--(\ref{eq:perp_cond_3}) 
with respect to $\cL(A\oplus{}B)\otimes\Omega$, 
and hence $\alpha(\cL(A\oplus{}B)\otimes\Omega)\ge2$. 
However, since $\alpha(\Omega)=1$, 
it follows from Proposition~\ref{Prop:fullyNoisy} that $\alpha(\cL(A\oplus{}B)\otimes\Omega)=1$, 
which leads to a contradiction. 
Thus, $(\cL(B,A)\otimes\Omega)^{\perp}$ has no rank-one operator. 
It follows that 
\begin{eqnarray}
\alpha(\Sigma\otimes\Omega)
&=&\max\{\alpha(S_{AA}\otimes\Omega), \alpha(T_{BB}\otimes\Omega)\} \\
&=&\max\{\alpha(S_{AA}), \alpha(T_{BB})\}\cdot\alpha(\Omega) \\
&=&\alpha(\Sigma)\alpha(\Omega), 
\end{eqnarray}
where the first equality follows from Corollary~\ref{cor:offnorank1}, 
the second equality is from the multiplicativity assumption for $\alpha$, 
and the last equality follows from 
the assumption $U_{BA}=\cL(B,A)$ and Corollary~\ref{cor:offnorank1}. 
\end{proof}

The following example shows two quantum channels 
whose noncommutative graphs satisfy 
the condition in Eq.~(\ref{eq:fulloff-0Omega}) of Theorem~\ref{Thm:add_BNG}. 
\begin{Example}
Let us consider a quantum channel $\cN: \cL(A\oplus{}B)\rightarrow\cL(C)$, 
where $A=B=C=\Complex^{2}$, 
with Kraus operators 
\begin{equation}
\begin{split}
\frac{1}{\sqrt2}\left(\ket{0}_{C}\bra{0}_{A}+\ket{1}_{C}\bra{1}_{A}\right), 
\frac{1}{\sqrt2}\left(\ket{0}_{C}\bra{1}_{A}+\ket{1}_{C}\bra{0}_{A}\right), \\
\frac{1}{\sqrt2}\left(\ket{0}_{C}\bra{0}_{B}+\ket{1}_{C}\bra{1}_{B}\right), 
\frac{1}{\sqrt2}\left(\ket{0}_{C}\bra{0}_{B}-\ket{1}_{C}\bra{1}_{B}\right), 
\end{split}
\end{equation}
and a quantum channel $\cM: \Complex^{3}\rightarrow\Complex^{3}$ 
with Kraus operators 
\begin{eqnarray}
\frac{1}{\sqrt2}(\ketbra{0}{0}+\ketbra{1}{1}), 
\frac{1}{\sqrt2}\ketbra{2}{2}, \\
\frac{1}{\sqrt2}(\ketbra{0}{1}-\ketbra{1}{2}), 
\frac{1}{\sqrt2}\ketbra{2}{0}. 
\end{eqnarray}
We can show that 
\begin{equation}
\begin{split}
S(\cN)=&\,\mathrm{span}\{\ketbra{+}{+},\ketbra{-}{-}\}_{AA} 
\oplus{}\mathrm{span}\{\ketbra{0}{0},\ketbra{1}{1}\}_{BB} \\
{}&\oplus{}\cL(B,A)\oplus{}\cL(A,B), \\
S(\cM)=&\,\mathrm{span}\{\ketbra{0}{1}+\ketbra{1}{2}, \ketbra{1}{0}+\ketbra{2}{1}\}^{\perp}, 
\end{split}
\end{equation}
where $\ket{\pm}=(\ket{0}\pm\ket{1})/{\sqrt2}$. 
It follows from Theorem~\ref{Thm:additivity_qubit} that 
the multiplicativity condition of Theorem~\ref{Thm:add_BNG} is satisfied. 
It is not hard to show that $S(\cM)^{\perp}$ has no rank-one operator. 
Thus, $\alpha(S(\cM))=1$ by Proposition~\ref{prop:alpha=1}, 
and then the condition in Eq.~(\ref{eq:fulloff-0Omega}) holds. 
Therefore, we conclude from Theorem~\ref{Thm:add_BNG} that 
$\alpha$ is multiplicative for $S(\cN)$ and $S(\cM)$, 
that is, $\cC_{0}^{(1)}$ is additive for the two channels $\cN$ and $\cM$. 
\end{Example}

We can derive the following corollary 
from the condition in Eq.~(\ref{eq:sumcap}) of Theorem~\ref{Thm:add_BNG}. 

\begin{Cor}\label{cor:add_BNG_1}
Let 
\begin{equation}
\Sigma=S_{AA}\oplus{}T_{BB}\oplus{}U_{BA}\oplus{}U^{\dag}_{AB}\subseteq\cL(A\oplus{}B) 
\end{equation}
is a block noncommutative graph such that $\alpha(S_{AA})=\alpha(T_{BB})=1$, 
and let $\Omega$ be a noncommutative graph. 
Suppose that $\alpha$ is multiplicative for $S_{AA}$ and $\Omega$, 
and also for $T_{BB}$ and $\Omega$. 
If $U^{\perp}_{BA}$ has a rank-one operator, 
then $\alpha$ is multiplicative for $\Sigma$ and $\Omega$. 
In particular, if $\dim(U_{BA})<\dim(A)+\dim(B)-1$, 
then $\alpha$ is multiplicative for $\Sigma$ and $\Omega$. 
\end{Cor}
\begin{proof}
Since $U^{\perp}_{BA}$ has a rank-one operator, 
there exist $\ket{\psi}$ and $\ket{\phi}$ such that 
$\ketbra{\psi}{\phi}\in{}U^{\perp}_{BA}$. 
Thus, $\ket{\psi}$ and $\ket{\phi}$ satisfy Eqs.~(\ref{eq:perp_cond_1})--(\ref{eq:perp_cond_3}), 
and so $\alpha(\Sigma)\ge2$ by Theorem~\ref{Thm:blockNGcap}. 
In other hands, it follows from Corollary~\ref{cor:offnorank1} that 
$\alpha(\Sigma)\le\alpha(S_{AA})+\alpha(T_{BB})=2$. 
Thus, 
\begin{equation}
\alpha(\Sigma)=\alpha(S_{AA})+\alpha(T_{BB}), 
\end{equation}
that is, the condition in Eq.~(\ref{eq:sumcap}) holds. 
It follows form Theorem~\ref{Thm:add_BNG} that 
$\cC_{0}^{(1)}$ is additive for $\Sigma$ and $\Omega$. 

Let us note that 
if $U^{\perp}_{BA}$ has no rank-one operator, 
in other words, any nonzero element of $U^{\perp}_{BA}$ has rank at least $2$, 
then the dimension of $U^{\perp}_{BA}$ is bounded above~\cite{CMW08} by 
\begin{equation}
\dim(U^{\perp}_{BA})\le(\dim(A)-1)(\dim(B)-1). 
\end{equation}
Thus, $U^{\perp}_{BA}$ has a rank-one operator 
when $\dim(U^{\perp}_{BA})>(\dim(A)-1)(\dim(B)-1)$, 
that is, $\dim(U_{BA})<\dim(A)+\dim(B)-1$. 
\end{proof}

\begin{Example}
Let us consider a quantum channel $\cN: \cL(A\oplus{}B)\rightarrow\cL(C\oplus{}D)$, 
where $A=C=D=\Complex^{4}$ and $B=\Complex^{2}$, 
with Kraus operators 
\begin{eqnarray}
E&=&\frac{3}{5}\sum_{i=0}^{3}\ket{i}_{C}\bra{i}_{A}, \\
F_{1}&=&\frac{1}{2}(\ket{0}_{C}\bra{0}_{B}+\ket{2}_{C}\bra{1}_{B}), \\
F_{2}&=&\frac{1}{2}(\ket{0}_{C}\bra{0}_{B}-\ket{2}_{C}\bra{1}_{B}), \\
F_{3}&=&\frac{1}{2}(\ket{1}_{C}\bra{0}_{B}+\ket{3}_{C}\bra{1}_{B}), \\
F_{4}&=&\frac{1}{2}(\ket{3}_{C}\bra{0}_{B}+\ket{1}_{C}\bra{1}_{B}), \\
G_{jk}&=&\frac{1}{5}\sum_{m=0}^{3}\omega^{mk}\ket{m\oplus{}j}_{D}\bra{m}_{A},
~j,k\in\{0,1,2,3\}, \qquad
\end{eqnarray}
where $\omega=e^{2\pi\mathrm{i}/4}$ is a $4$th root of unity 
and $\oplus$ denotes addition modulo 2. 
It is straightforward to show that 
\begin{equation}
S(\cN)=\cL(A)\oplus{}\mathrm{span}\{I_{2},X,Z\}_{BB}\oplus{}U_{BA}\oplus{}U^{\dag}_{AB}, 
\end{equation}
where $X=\ketbra{0}{1}+\ketbra{1}{0}$, $Z=\ketbra{0}{0}-\ketbra{1}{1}$, 
and $U_{BA}=\mathrm{span}\{F_{s} : s=1,2,3,4\}$. 
It is easy to show that 
$\cL(A)^{\perp}$ and $\mathrm{span}\{I_{2},X,Z\}_{BB}^{\perp}$ have no rank-one operator. 
Thus, we have 
\begin{equation}
\alpha(\cL(A))=\alpha(\mathrm{span}\{I_{2},X,Z\}_{BB})=1 
\end{equation}
from Proposition~\ref{prop:alpha=1}. 
It follows from Proposition~\ref{Prop:fullyNoisy} and Theorem~\ref{Thm:additivity_qubit} that 
the multiplicativity condition of Corollary~\ref{cor:add_BNG_1} holds 
for any noncommutative graph $\Omega$. 
Finally, we observe that 
\begin{equation}
\dim(U_{BA})\le4<5=\dim(A)+\dim(B)-1. 
\end{equation}
It follows from Corollary~\ref{cor:add_BNG_1} that 
$\alpha$ is multiplicative for $S(\cN)$ and any noncommutative graph $\Omega$, 
that is, $\cC_{0}^{(1)}$ is additive for the channel $\cN$ and any quantum channel $\cM$. 
\end{Example}

\section{summary}\label{sec:Sum}
In this paper, we have established a rigorous theoretical framework that identifies when the zero-error capacities for specific classes of quantum channels are additive by investigating the multiplicativity of the independence number of noncommutative graphs induced by the quantum channels. We have first presented some sufficient conditions under which the independence number is multiplicative, ensuring the additivity of the one-shot zero-error classical capacity. These conditions have been shown to extend to the asymptotic zero-error classical capacity, providing a broader scope for our findings. Furthermore, we have investigate the structural properties of noncommutative graphs in block form to derive specific criteria for multiplicativity within this architecture. Finally, we have provided explicit examples of quantum channels to demonstrate the practical application of our additivity conditions and validate the theoretical results.

When we are given noncommutative graphs $S_1$, $S_2$, and $T$, 
where $\cC_{0}^{(1)}$ is additive for $S_1$ and $T$, and also for $S_2$ and $T$, Theorem~\ref{Thm:add_BNG} and Corollary~\ref{cor:add_BNG_1} provide a method to construct a new noncommutative graph $S$ from $S_1$ and $S_2$ 
such that $\cC_{0}^{(1)}$ is additive for $S$ and $T$. 
For noncommutative graphs $\Sigma$ and $\Omega$ satisfying the conditions 
in Theorem~\ref{Thm:add_BNG} or Corollary~\ref{cor:add_BNG_1}, 
their associated quantum channels exist as stated in Remark~\ref{Rem:ch-NG}. 
However, it remains unclear which quantum channels correspond precisely to a given block noncommutative graph $\Sigma$. More generally, this raises the problem of characterizing the classes of quantum channels corresponding to a given block noncommutative graph. These questions point to promising directions for future research in the field.

Although we have considered the direct sum of two Hilbert spaces and the associated noncommutative graphs in $2 \times 2$ block form in Section~\ref{sec:Block}, we note that this framework admits a natural generalization to multi-block forms, that is, 
\begin{equation}
\Gamma\equiv\bigoplus_{i,j=1}^{d}S_{ij}\subseteq\cL\left(\bigoplus_{k=1}^{d}A_k\right) 
\end{equation}
is a noncommutative graph, 
where $A_k$ are Hilbert spaces,  and $S_{ij}$ are subspaces of $\cL(A_i,A_j)$ such that 
$S_{ij}^{\dag}=S_{ji}$ and $I_{A_i}\in{}S_{ii}$. 

Moreover, an inductive application of Theorem~\ref{Thm:blockNGcap} can also provide us with the condition for the independent number of $\Gamma$;
$\alpha(\Gamma)$ is the maximum number of 
\begin{equation}
\sum_{k=1}^{d}|\{\ket{v^{k}_{i_k}}\in{}A_k : i_k=1,\dots,m_k\}|, 
\end{equation}
where all $\ket{v^{k}_{i_k}}$'s are nonzero vectors such that 
for all $k\in\{1,\dots,d\}$, 
\begin{equation}
\ketbra{v^{k}_{i_k}}{v^{k}_{i_{k}'}}\perp{}S_{kk},
~\forall{}{i_k},{i_k'}\in\{1,\dots,m_k\}~\text{with}~{i_k}\ne{i_k'}, \qquad
\end{equation}
and for all $k,l\in\{1,\dots,d\}$ with $k\ne{}l$, 
\begin{equation}
\ketbra{v^{k}_{i_k}}{v^{l}_{i_l}}\perp{}S_{lk},
~\forall{i_k}\in\{1,\dots,m_k\},{i_l}\in\{1,\dots,m_l\}. \qquad
\end{equation}
Having established these relevant ingredients, a natural direction for future research is to generalize the additivity conditions in Theorem~\ref{Thm:add_BNG} and Corollary~\ref{cor:add_BNG_1} to multi-block noncommutative graphs.

\acknowledgments
This work was supported by 
the Institute for Information \& Communications Technology Planning \& Evaluation (IITP) 
grant funded by the Korean government (MSIP) (Grant No. RS-2025-02304540). 
JSK was supported by Creation of the Quantum Information Science R\&D Ecosystem 
(Grant No. 2022M3H3A106307411) through the National Research Foundation of Korea (NRF) 
funded by the Korean government (Ministry of Science and ICT).



\begin{thebibliography}{9}

\bibitem{Shan56}
C.~E.~Shannon, 
The zero error capacity of a noisy channel, 
IRE Trans. Inf. Theory \textbf{2}, 8 (1956). 

\bibitem{KO98}
J.~K\"{o}rner and A.~Orlitsky. 
Zero-error information theory. 
IEEE Trans. Inf. Theory \textbf{44}, 2207 (1998). 

\bibitem{MA05}
Rex~A.~C.~Medeiros and Francisco~M.~De~Assis, 
Quantum Zero-error Capacity, 
Int. J. Quantum Inf. \textbf{3}, 135 (2005). 

\bibitem{Duan09}
R.~Duan, 
Super-Activation of Zero-Error Capacity of Noisy Quantum Channels, 
arXiv:0906.2527. 

\bibitem{Shirokov15}
M.~E.~Shirokov, 
On channels with positive quantum zero-error capacity having vanishing n-shot capacity, 
Quantum Inf Process \textbf{14}, 3057 (2015). 

\bibitem{LY16}
D.~Leung and  N.~Yu, 
Maximum privacy without coherence, zero-error,
J. Math. Phys. \textbf{57}, 092202 (2016). 

\bibitem{CLMW11}
T.~S.~Cubitt and D.~Leung and W.~Matthews and A.~Winter, 
Zero-Error Channel Capacity and Simulation Assisted by Non-Local Correlations, 
IEEE Trans. Inf. Theory \textbf{57}, 5509 (2011). 

\bibitem{DSW13}
R.~Duan, S.~Severini, and A.~Winter. 
Zero-Error Communication via Quantum Channels, Noncommutative Graphs, and a Quantum Lovász Number. 
IEEE Trans. Inf. Theory \textbf{59}, 1164 (2013). 

\bibitem{DW16}
R.~Duan and A.~Winter, 
No-Signalling-Assisted Zero-Error Capacity of Quantum Channels and an Information Theoretic Interpretation of the Lovász Number, 
IEEE Trans. Inf. Theory \textbf{62}, 891 (2016). 

\bibitem{BS07}
S.~Beigi and P.~W.~Shor, 
On the Complexity of Computing Zero-Error and Holevo Capacity of Quantum Channels, 
arXiv:0709.2090. 

\bibitem{CCH11}
T.~S.~Cubitt and J.~Chen and A.~W.~Harrow, 
Superactivation of the Asymptotic Zero-Error Classical Capacity of a Quantum Channel, 
IEEE Trans. Inf. Theory \textbf{57}, 8114 (2011). 

\bibitem{PH18}
J.~Park and J.~Heo, 
Activation of zero-error classical capacity in low-dimensional quantum systems, 
Phys. Rev. A \textbf{97}, 062314 (2018). 

\bibitem{CMW08}
T.~S.~Cubitt and A.~Montanaro and A.~Winter, 
On the dimension of subspaces with bounded Schmidt rank, 
J. Math. Phys. \textbf{49}, 022107 (2008). 

\end{thebibliography}
\end{document}